# Carpet cloaking on a dielectric half-space

Pu Zhang,[1,2] Michaël Lobet,[2] and Sailing He[1,2*]

[1] *JORCE (KTH-ZJU Joint Center of Photonics), Centre for Optical and Electromagnetic Research, East Building # 5, Zijingang campus, Zhejiang University (ZJU), Hangzhou 310058, China*
[2] *Department of Electromagnetic Engineering, School of Electrical Engineering, Royal Institute of Technology (KTH), S-100 44 Stockholm, Sweden*
*\* Corresponding author: sailing@kth.se*

**Abstract:** Carpet cloaking is proposed to hide an object on a dielectric half-space from electromagnetic (EM) detection. A two-dimensional conformal transformation specified by an analytic function is utilized for the design. Only one nonsingular material parameter distribution suffices for the characterization. The cloaking cover situates on the dielectric half-space, and consists of a lossless upper part for EM wave redirection and an absorbing bottom layer for inducing correct reflection coefficient and absorbing transmission. Numerical simulations with Gaussian beam incidence are performed for verification.

**1. Introduction**

Since the introduction of a perfect electromagnetic (EM) cloak based on transformation optics [1], this field has been developing rapidly. Besides cloaks, many other novel devices have also been designed using transformation optics [2-4]. However, perfect cloaks require singular material parameters. To overcome this difficulty in realization, much effort towards simplified models [5-7], experiments [8-11] and other invisibility mechanisms [12-14] has been made. Among these progresses, carpet cloaking [15] is a good compromise between functionality and feasibility. It utilizes a nonsingular mapping to transform a perfect electrical conductor (PEC) ground from flat to convex, making some space beneath for concealment. Outside observers above the PEC ground can not perceive the existence of the bump. Singularity in material parameters is avoided by using the nonsingular transformation. Furthermore, the application of (quasi-) conformal mapping justifies the one-parameter characterization of the cloak. Different from perfect cloaks, only half-space cloaking above a PEC ground was achieved. In fact, half-space cloaking above a dielectric half-space was studied even earlier [16], where two matching strips are buried under a semi-cylindrical cloaking cover to produce the same reflection as that from the bare half-space. Very recently, special interest is greatly aroused by the experimental realizations of carpet cloaks at both microwave [9] and optical [10,11] frequencies.

The original theory for carpet cloaking is for invisibility on a PEC ground, and all subsequent works still stick to the same assumption. In practical applications, however, a dielectric (especially with loss) ground is much more common and practical than a PEC ground. In this paper, carpet cloaking is extended to the dielectric ground situation, so that the reflection from the carpet cloak is the same as that from a bare dielectric ground. In the configuration (see Fig. 1), the concealment volume is also created by transforming the dielectric interface from flat to convex with a nonsingular mapping. An additional absorbing layer is introduced, instead of just a PEC boundary, to produce correct reflection coefficient and prevent EM wave from penetrating into the concealment volume. The design is elaborated in Section 2, followed by the numerical verification and discussions in Section 3 and conclusion in Section 4.

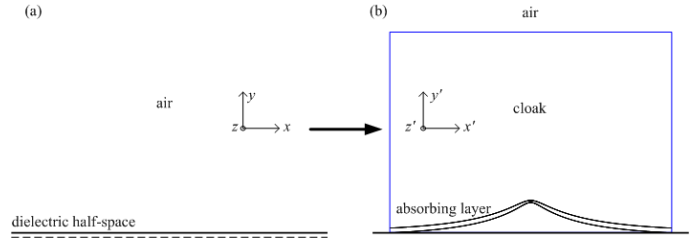

Fig. 1. Schematic diagram for carpet cloaking on a dielectric half-space: Air and the bare dielectric half-space in virtual space (a) are transformed to the carpet cloak in the physical space (b).

**2. Design of the carpet cloak on a dielectric half-space**

To design an invisibility cloak with transformation optics, we always start from a virtual space. Here we consider a virtual space with a dielectric half-space as shown in Fig. 1(a). The dielectric half-space is characterized by a complex permittivity of $\varepsilon_g = \varepsilon_r - \varepsilon_i * i$ (harmonic time dependence $\exp(i\omega t)$ is assumed and suppressed throughout the paper). For half-space cloaking, only EM waves above the half-space concern us. Refracted field is effectively attenuated in the lossy half-space within a finite layer indicated by the dashed line in Fig. 1(a). This layer, together with air space above the half-space, constitutes the region to be transformed, under which nearly no refracted EM field exists. After the transformation, the layer becomes curved, forming some space for concealment in the physical space. In practice, the cloak should have a finite volume as enclosed by the blue line in Fig. 1(b). Therefore, outside this volume the transformation must be very close (if not equal) to identity so that the material parameters can be approximated by air and the reflected field remains nearly the same as that from the bare dielectric half-space. The configuration is assumed to be two-dimensional (2D) and transverse electrically (TE, $E_z$-) polarized for simplicity.

Conformal mapping is preferred in transformation optics designs, since the resulted material can be described by only one parameter for a specified polarization, and consequently much easier to realize. Real and imaginary parts of analytical functions of a complex variable are naturally used to introduce conformal mappings [17,18]. Here we use a simple analytical function of $z' = z - 1/z$, where $z = x + y * i$ and $z' = x' + y' * i$ represent 2D coordinate variables in the virtual and physical spaces, respectively. This function fulfills all the required properties mentioned above. $z'$ approaches to $z$ when $z$ goes to infinity. The layer under the interface bounded by $y = 1.3$ and $y = 1.4$ (arbitrary unit is used throughout the paper) is transformed to the curved absorbing layer shown in Fig. 1(b). Expressing explicitly the real and imaginary parts of the function in terms of the spatial coordinates, we obtain the following transformation

$$x' = x\left(1 - \frac{1}{x^2 + y^2}\right), \quad y' = y\left(1 + \frac{1}{x^2 + y^2}\right), \quad z' = z. \tag{1}$$

The inverse transformation will be used during the calculation of the material parameter. It can be easily derived as

$$x = \frac{x' + p}{2}, \quad y = \frac{y' + q}{2}, \quad z = z', \tag{2}$$

where $p = \text{sgn}(x')\sqrt{(s + \sqrt{s^2 + t^2})/2}$, $q = \sqrt{(-s + \sqrt{s^2 + t^2})/2}$, $s = x'^2 - y'^2 + 4$, and $t = 2x'y'$. According to transformation optics, the material parameter tensor of the carpet cloak in the physical space can be calculated as $\overline{\overline{J}}\overline{\overline{J}}^T/\det(\overline{\overline{J}})$ using Jacobian matrix $\overline{\overline{J}} = (\partial x'^i/\partial x^j)$ of the transformation. For the special conformal transformation, the tensor reduces to a diagonal one with the first two components being 1 and the $zz$-component nontrivial. Using the transformation in Eq. (1), the material parameter for the carpet cloak in the physical space becomes

$$\eta \equiv \frac{\varepsilon_l}{\varepsilon_g} = \frac{\varepsilon_c}{\varepsilon_0} = \left[1 + \frac{2(x^2 - y^2) + 1}{(x^2 + y^2)^2}\right]^{-1}, \tag{3}$$

where $\varepsilon_l$ and $\varepsilon_c$ are permittivities of the absorbing layer and the upper part of the carpet cloak, respectively. It is noted that Eq. (3) is formulated with coordinate variables in the virtual space, and the inverse transformation in Eq. (2) must be applied during the implementation. Also note that ratio $\eta$ approaches 1 when $|z|$ becomes very large. Thus the cloak can be approximately confined in a finite region. In the present design, the vertical boundaries are fixed at positions where the absorbing layer touches the half-space, and the upper boundary is at $y' = 5.8$. Numerical evaluation of $\eta$ in the carpet cloak shows that it can range from 0.85 to 6.00. At the interface between the carpet cloak and air, the deviation of $\eta$ from 1 is within about 0.1, so the approximation of the truncation is good enough. Little additional reflection is expected at the interface between the carpet cloak and air.

Up till now, the dielectric half-space is assumed to be lossy, and thus the refracted EM wave is sufficiently attenuated inside the absorbing layer. On the other hand, if the dielectric half-space is lossless or has very low loss, the absorbing layer can not be designed by using the previous procedure directly. However, the carpet cloak can still be introduced in a similar way. In this case, fictitious loss is added to the dielectric half-space in the virtual space to attenuate the refracted EM wave. Additional reflection may be caused by the fictitious loss adversely.

In order to keep the additional reflection to a negligible level, we use a gradient loss of smooth function (e.g., quadratic) with the loss increasing gradually from 0 at the interface of the dielectric half-space. Now the refracted EM wave can be effectively attenuated within a finite layer as before. Then the absorbing layer of the carpet cloak is constructed by applying the transformation in the same way as before.

## 3. Numerical verification and discussions

In order to validate and visualize the present design of carpet cloaking on a dielectric half-space, numerical simulations based on the finite element method are carried out. The configuration for simulations is the same as that shown in Fig. 1. Perfectly matched layers are used to terminate the computational domain. An obliquely incident TE polarized Gaussian beam can be used as an excitation. Free space wavelength is set to 0.12. Two different dielectric half-spaces are considered in the following simulations. In the first example, the dielectric half-space is described by $\varepsilon_g / \varepsilon_0 = 5 - 5i$ (a typical permittivity for soils with certain moisture at microwave frequencies [19]). The material parameters for the upper part and the absorbing layer of the carpet cloak are then evaluated with Eq. (3). For the convenience of comparison, three different settings are simulated with the same excitation and snapshots of the electric field distributions are shown in Fig. 2. In Fig. 2(a), a PEC bump is placed on the dielectric half-space, and the incident Gaussian beam is scattered violently over a large range of direction. Then the carpet cloak is applied to hide the PEC bump in Fig. 2(b), where the reflected field is restored to a Gaussian beam to an expected direction. The refracted wave is efficiently absorbed within the absorbing layer, and no field reaches the PEC bump. Fig. 2(c) presents the reflection of the incident Gaussian beam from the bare dielectric half-space. By comparison, one sees the restored Gaussian beam in Fig. 2(b) is of the same direction and intensity as that in Fig. 2(c).

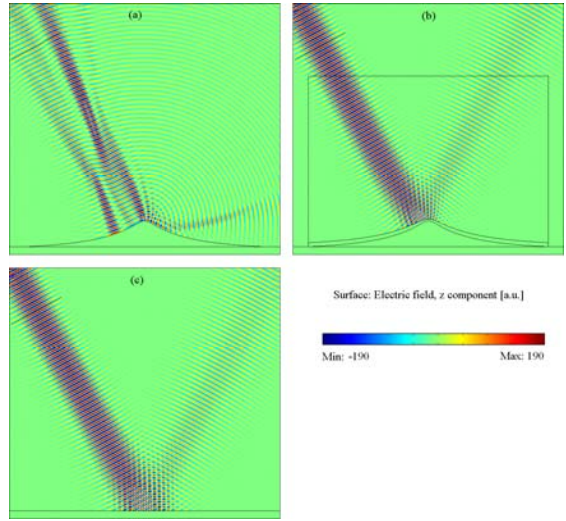

Fig. 2. Snapshots of the electric field distributions for a Gaussian beam impinging upon (a) a PEC bump on a dielectric half-space characterized by $\varepsilon_g / \varepsilon_0 = 5 - 5i$, (b) the PEC bump covered by a carpet cloak, and (c) the bare dielectric half-space.

In the second simulation example, a lossless dielectric half-space with $\varepsilon_g / \varepsilon_0 = 2$ is considered. As elucidated in the previous section, the absorbing layer of the carpet cloak is now designed by adding some fictitious gradient loss. The permittivity is modified to a complex one, $\varepsilon_g [1 - 150 (1.4 - y)^2 i]$ within the layer of $1.3 \leq y \leq 1.4$ (as before), which is then transformed to obtain the absorbing layer in the physical space. Our simulation shows that this strategy works well. The relative permittivity distribution of the carpet cloak is implemented by using Eq. (3), ranging from 0.87 to 12.00. When a background material other than air [9-11] is considered, the range can be scaled up to above 1, making it realizable at optical frequencies with air diluted $SiO_2$ and Si structures [10,11]. Broadband behavior can also be expected due to the low dispersion. All the other simulation settings are the same as those in the previous simulation example. Simulation results are shown in Fig. 3. The carpet cloak again restores the reflection of the incident Gaussian beam. Refracted EM wave is effectively attenuated in the absorbing layer. No additional reflection is observed. Comparison of Fig. 3(a) and Fig. 3(b) reveals that the reflected beams have the same amplitude and direction, rendering the PEC bump invisible to external observers.

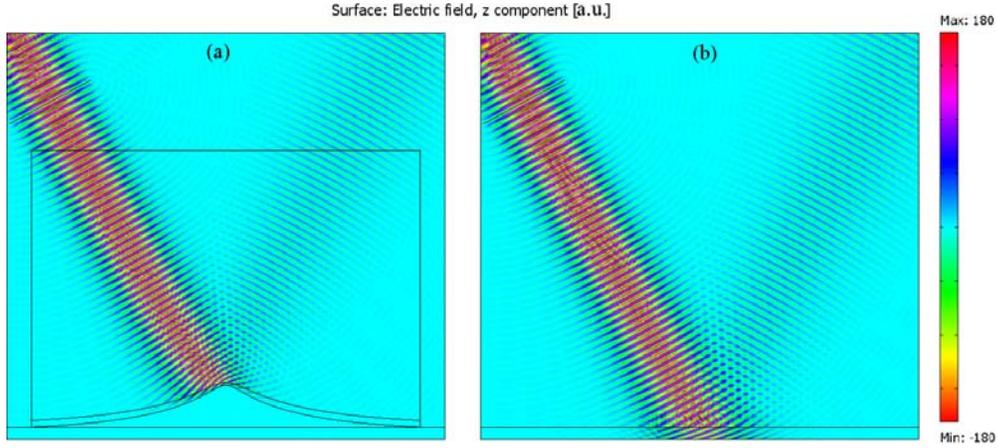

Fig. 3. Snapshots of the electric field distributions for a Gaussian beam impinging upon (a) a carpet cloak applied to a PEC bump on a lossless dielectric half-space ($\varepsilon_g / \varepsilon_0 = 2$), and (b) the bare dielectric half-space.

Verified by the two simulation examples, the proposed carpet cloak can make objects on a dielectric half-space invisible. However, a close observation of Fig. 2 and Fig. 3 discloses a minor shift of the reflected Gaussian beam from the carpet cloak with respect to that from the bare dielectric half-space. This imperfection can be explained as the effect of the truncation of the conformal mapping. A further reduction of this shift can be expected by optimally selecting analytical functions, which is not included in the present paper.

### 4. Conclusion

In summary, carpet cloaking on a dielectric half-space has been proposed in this paper. As an extension of earlier carpet cloaking on a PEC ground, the present design is applicable for more extensive situations. Similar to the conventional carpet cloak, the planar interface of the dielectric half-space in a virtual space is transformed to convex by a conformal mapping, forming a concealment volume. Here, a conformal mapping specified by an analytical complex function has been utilized for simplicity and analyticity. For both lossy and lossless dielectric half-spaces, an absorbing layer is introduced at the bottom of the carpet cloak to give correct reflection coefficient and to attenuate the refracted EM wave. The carpet cloak is characterized by only one nonsingular material parameter and thus broadband cloaking is feasible experimentally at both microwave and optical frequencies. Numerical simulations based on the finite element method have been performed to verify the effectiveness of the present carpet cloak.

**Acknowledgments**

This work is partially supported by the National Natural Science Foundation (Grants No. 60990320) of China, the Swedish Research Council (VR) and AOARD. Michaël Lobet is also supported by an EU scholarship as an Erasmus Mundus exchange student from Research Center in Physics of Matter and Radiation (PMR), University of Namur (FUNDP), Belgium.